\documentclass[aps,prd,showpacs,preprintnumbers,twelvepoint,floatfix]{revtex4}
\usepackage{epsfig}
\usepackage{latexsym,amsmath,amssymb,amstext}
\usepackage{float}
\usepackage{graphicx}
\usepackage{color}

\newcommand\bef{\begin{figure}}
\newcommand\eef[1]{\label{fg:#1}\end{figure}}
\newcommand\beq{\begin{equation}}
\newcommand\eeq[1]{\label{#1}\end{equation}}
\newcommand\beqa{\begin{eqnarray}}
\newcommand\eeqa[1]{\label{#1}\end{eqnarray}}
\newcommand\bet{\begin{table}}
\newcommand\eet[1]{\label{tb:#1}\end{table}}

\newcommand\fgn[1]{Figure \ref{fg:#1}}
\newcommand\eqn[1]{eq.\ (\ref{#1})}

\newcommand\ie{{\sl i.e.\/}}

\newcommand\etal{{\sl et al.\/}}
\newcommand\ibid{{\sl ibid.\/}}
\newcommand\jhep{{\sl J.\ H.\ E.\ P.\/}\ }

\newcommand\npa{{\sl Nucl.\ Phys.\/}\ A}
\newcommand\npb{{\sl Nucl.\ Phys.\/}\ B}

\newcommand\plb{{\sl Phys.\ Lett.\/}\ B}

\newcommand\alphas{\ensuremath{\alpha_{\scriptscriptstyle S}}}
 
\newcommand\jpsi{\ensuremath{J/\psi}}

\newcommand{\pbpb}{{\rm{\scriptscriptstyle PbPb}}\/}
\newcommand{\M}{{\cal M}}

\newcommand{\T}{{\scriptscriptstyle T}}

\begin{document}

\title{On the thermalization of quarkonia at the LHC}

\author{Sourendu\ \surname{Gupta}}
\email{sgupta@theory.tifr.res.in}
\affiliation{Department of Theoretical Physics, Tata Institute of Fundamental
         Research,\\ Homi Bhabha Road, Mumbai 400005, India.}
\author{Rishi\ \surname{Sharma}}
\email{rishi@theory.tifr.res.in}
\affiliation{Department of Theoretical Physics, Tata Institute of Fundamental
         Research,\\ Homi Bhabha Road, Mumbai 400005, India.}

\begin{abstract}
We argue that the relative yields of $\Upsilon$ states observed at the
LHC can be understood as bottomonium states coming to early thermal
equilibrium and then freezing out. The bottomonium freezeout temperature
is approximately 250 MeV.  We examine its systematics as a function of
centrality. We remark on the interesting differences seen by the CMS
and ALICE experiments in the charmonium sector.
\end{abstract}

\pacs{12.38.Mh, 11.15.Ha, 12.38.Gc}
\preprint{TIFR/TH/14-03}
\maketitle

A thermal medium can disrupt the formation of a bound state of a heavy
quark, $Q$ and antiquark, $\bar Q$ \cite{matsui}.  This is a multi-scale
problem, involving the temperature $T$ and the quark mass $M$. The quark
is heavy, \ie, $M/T\gg1$.  For the charmonium this ratio is about 5--10
and for the bottomonium it is around 15--30; so both these flavours can
be considered to be heavy in this sense. Also, $M/\Lambda\gg1$, where
$\Lambda\simeq m_\rho/2$ is the typical scale of QCD \footnote{This can
be thought of as either an effective constituent light quark mass, or as
the scale which controls the running of the strong coupling, \alphas. The
results of dimensional analysis do not depend on this choice. We will
use the notation $X\simeq Y$ to mean that the quantities $X$ and $Y$
are of similar order, in the sense of dimensional analysis.}. This
second comparison implies that in the $\bar QQ$ bound state the quarks
are slow, with velocity $v^2\simeq0.23$ for charm and $v^2\simeq0.08$
for bottom \cite{tye}. In NRQCD counting, the binding energy $B\simeq
Mv^2\simeq\Lambda$ in both cases \cite{bbl}. For the temperature range
of relevance, we find that $B/T\simeq\Lambda/T \simeq1$.  As a result,
thermal effects can drastically modify the bound state.

Since a thermal medium can be formed in PbPb collisions, but not in pp
collisions, there should be certain systematic differences between
the yields in these two cases \cite{pattern}. Such effects were first
seen at the SPS \cite{sps}, but there were important backgrounds to
the signal which came from the initial state via parton density effects
\cite{satz} and the final state through comover interactions \cite{other}.
Furthermore, the observed effects switched on slowly with centrality
and nuclear size, and so were difficult to interpret clearly \cite{gavai}.

At the LHC the experimental situation has changed drastically. The
the 1S, 2S and 3S states of the $\Upsilon$ \cite{lhc-cms-upsilon} and
the 1S and 2S states of the $J/\psi$ \cite{lhc-cms-psi,lhc-alice-psi}
have been studied, and clear differences between pp and PbPb collisions
have been established. For any hadron $h$, we will use the notation
$R_{\pbpb}[h]=N_{\pbpb}[h]/N_{pp}[h]$ where $N_{\pbpb}[h]$ is the
yield of $h$ in PbPb collisions and $N_{pp}[h]$ in pp collisions.
The new experimental results show sequential suppression \cite{digal}
very clearly: $R_{\pbpb}[\Upsilon(1S)]$, $R_{\pbpb}[\Upsilon(2S)]$
and $R_{\pbpb}[\Upsilon(3S)]$ are not equal, nor are $R_{\pbpb}[\jpsi]$
and $R_{\pbpb}[\psi(2S)]$.

Initial state effects cannot be the explanation, since the relevant
values of the Bjorken variable are almost equal for the three bottomonium
states, and a different common value for the two charmonium states,
implying that parton density effects would be the same.  Final state
comover interactions were invoked even at the LHC in order to explain
the observed suppression of {\jpsi} \cite{lhc-comover}. The comoving
material which is thermalized gives rise to the signal. The comovers
responsible for the background are unthermalized and relatively cold
spectators from the initial PbPb collision. However, the data are taken
at central rapidity and separated from the spectator fragments by $\Delta
y\simeq6$. So comover interactions cannot be the explanation for the
observations. In this cleaner environment it would be interesting to
check whether the data allow a thermal interpretation.

We begin by noting that the equilibrium density of heavy quarkonia, with
mass $\M=2M-B\simeq M(2-v^2)$ and fugacity $z$,
\beq
   n\simeq (\M T)^{3/2}\exp\left(-\frac{\M}T\right) z,
\eeq{density}
is small. As a result, the mean distance between quarkonia in thermal
equilibrium, $\lambda$, is large. In fact, we find that $T\lambda =
z^{-1/3} \sqrt{T/\M} \exp(2\M/3T)$. For $z=1$ this dimensionless number is
over 100 for charmonia and over 10 million for bottomonia. So, for both
the charm and bottom systems, the exponential dominates, and $T$ is much
larger than the inverse of the mean inter-quarkonium spacing. However,
essentially all the $\bar QQ$ pairs in question come from initial hard
scatterings, so $z$ should be computed in perturbative QCD, and the values
are much larger than unity \cite{pattern}. Nevertheless, the hierarchy
of length scales in the fireball remains $M\gg B\simeq T\gg1/\lambda$.

This dimensional analysis indicates that the net production rate of
quarkonia in thermal equilibrium may be obtained largely by understanding
the changes in the spectral density of the quarkonium states in
thermal equilibrium. Current attempts at extracting them in lattice QCD
indicate that there is generally an increase in the width, $\Gamma$,
of these quarkonium states at finite $T$ \cite{lattice}.  This width
has an interpretation as the inclusive rate of all reactions in which
the quarkonium state goes into an unbound $\bar QQ$ pair.  In fact,
understanding the phenomenon of quarkonium yields does not require a
full knowledge of the spectral function; knowing the reaction rate,
$\Gamma$, suffices.  It was pointed out in \cite{thews} that one must
add the reverse recombination processes to the screening \cite{matsui}
and dissociation \cite{bhanot} previously considered.  In fact,
if the heavy-quarkonium system comes to thermal equilibrium, then detailed
balance requires both binding and unbinding reactions.

Defining the formation time, $\tau=1/\Gamma$, NRQCD power counting has
been used to argue that $1/\tau\simeq B=Mv^2$ \cite{strickland}.  This
implies that $1/\tau$ is of order $\Lambda$ and hence non-perturbative.
However, this estimate does not involve $T$. Some phenomenological
estimates can be found in \cite{rapp,rishi}. A more systematic approach
is to perform a weak coupling expansion in a static limit of NRQCD at
finite temperature \cite{laine,brambilla}, from which the estimates
$\Gamma\simeq g^2T$ are obtained.  However, to control weak coupling
estimates, one has to have $\alphas\ll1$ and hence $T\gg\Lambda$.
Since we estimated that $B\simeq\Lambda\simeq T$, these weak-coupling
estimates, while illuminating, may not be entirely relevant to the
physical situation. Nevertheless, if one boldly extends this estimate
to the region $g\simeq1$, then $\Gamma\simeq T$, a result which is
generic in effective models. This simple scaling is also consistent with
recent lattice computations which indicate that for both the $\eta_b$
and $\Upsilon(1S)$ one has $\Gamma=T$ (within statistical errors)
immediately above $T_c$.  \cite{aarts}.

If $\Gamma$ for each of the resonances is large, then this would imply
that the various bound and unbound levels of the $\bar QQ$ system could
come to thermal equilibrium with each other in a very short time. Such
a situation is far simpler than that in $pp$ collisions, where one
must take into account details of how primary produced $\bar QQ$ pairs
hadronize into mesons. A thermalized system of this kind would stay
in equilibrium until $\Gamma$ falls below the expansion rate and the
heavy-quark system freezes out. This resembles models of statistical
hadronization \cite{thermal}. However, unlike those, we do not force
the heavy-quarkonia to freeze out at the same temperature as particles
made with lighter quarks. Instead we let the data decide the freezeout
temperature.

\bef
\begin{center}
\includegraphics[scale=0.7]{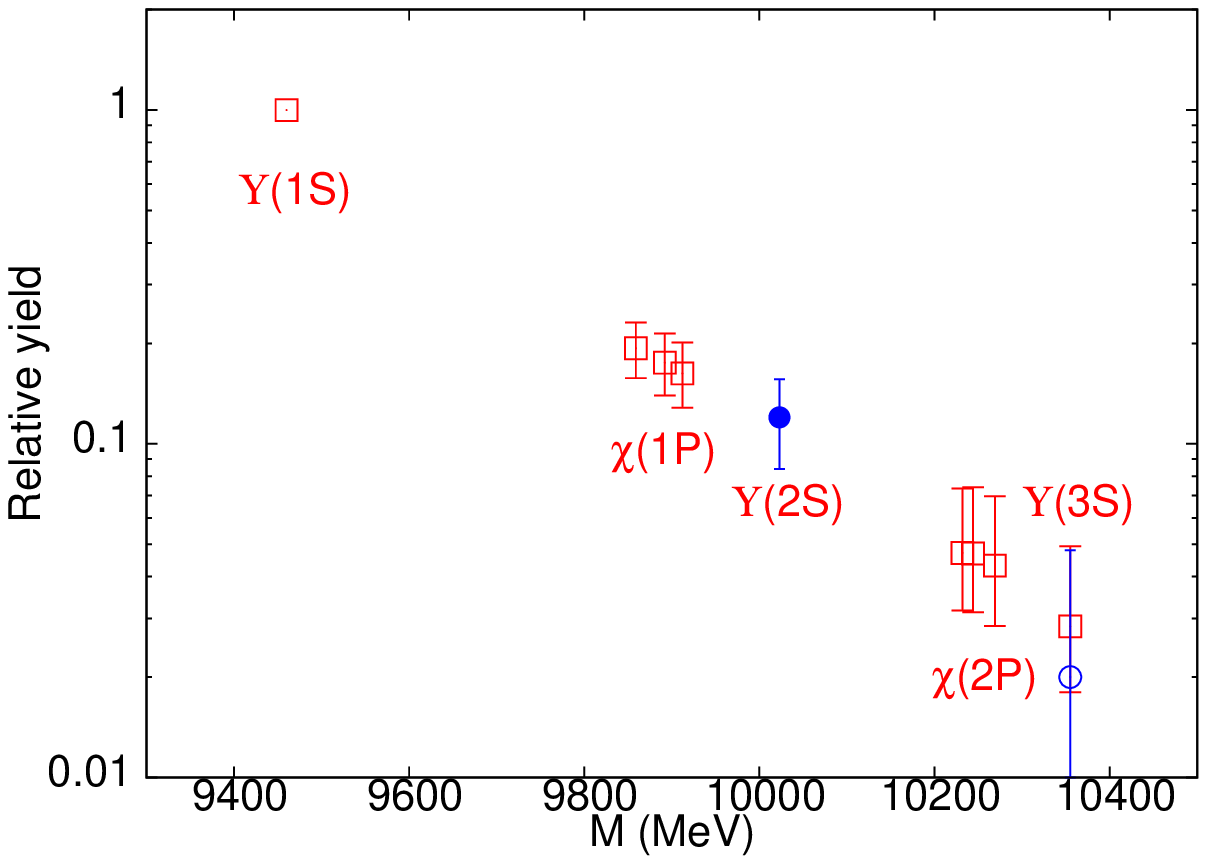}
\end{center}
\caption{Predictions for the yields of bottomonium mesons relative to the
 $\Upsilon(1S)$, using the simple model described in the text. The filled
 circle, corresponding to the relative yield of the $\Upsilon(2S)$ state,
 is used as an input. The rest are predictions. The observed data for the
 $\Upsilon(3S)$ state is shown as an unfilled circle.}
\eef{bottom}

While the ratio $R_{\pbpb}$ is very useful for comparing pp and PbPb collisions,
it is not the most appropriate quantity for examining thermal physics. Instead
we examine the relative yields of different resonances. The ratios of
the yields of various $\Upsilon$ states to that of the $\Upsilon(1S)$
are quoted in \cite{lhc-cms-upsilon}---
\beqa
 \nonumber
  Y[\Upsilon(2S)]
     &=& \frac{N_{\pbpb}[\Upsilon(2S)]}{N_{\pbpb}[\Upsilon(1S)]}
     = 0.12 \pm0.03 \pm0.02 \\
  Y[\Upsilon(3S)]
     &=& \frac{N_{\pbpb}[\Upsilon(3S)]}{N_{\pbpb}[\Upsilon(1S)]}
     = 0.02 \pm0.02 \pm0.02 
\eeqa{cms-values}
The bottomonium states were reconstructed from data taken at beam
energy of 2.76 TeV per nucleon from the muons they decay into. The muon
transverse momenta were greater than 4 GeV and rapidity less than 2.4.
The large error bar on the second ratio means that it cannot be used as
an input in the subsequent computations.  

We ask whether these two pieces of data can be explained by a thermal
model such as in \eqn{density}.  Fortunately, when we consider the yield
ratios $Y$, there is just a single parameter in this model, namely the
freezeout temperature, $T_f$.  However, one cannot just insert the values
in \eqn{cms-values} into \eqn{density}, and invert it to obtain the
temperature. One must take into account the fact that after the mesons
freeze out, the excited states may decay into the lower-lying quarkonium
states before reaching the detector. This modifies the primordial
densities. This has to be corrected for when extracting $T_f$. It turns
out that inverting \eqn{density} gives a reasonably close approximation,
and an efficient method for extracting $T_f$ is an iterative improvement
over this guess.  Adding the errors in \eqn{cms-values} in quadrature,
and neglecting the errors in the branching ratios, we find
\beq
   T_f=252^{+40}_{-39} {\rm\ MeV\/}
\eeq{param} 
from the first ratio, and $228^{+57}_{-228}$ MeV from the second. These
two estimates are compatible with each other, so we can use the first
number in our extraction.

This single parameter now gives predictions for the yields of the other
bottomonium mesons. These are shown in \fgn{bottom}.  The compatibility
of the yield of the $\Upsilon(3S)$ with the 1S and 2S states is visible
in this figure as the agreement between the prediction, based on
$Y[\Upsilon(2S)]$, and the direct measurement.  The predictions for the
$\chi$ states is made under the assumption that they freeze out at the
same temperature. It is generally seen in lattice computations that the P
states have significantly higher widths than the S states \cite{lattice}.
If so, then they would freeze out later. The relative yields shown in
\fgn{bottom} are therefore upper limits of expectations.

\bef
\begin{center}
\includegraphics[scale=0.7]{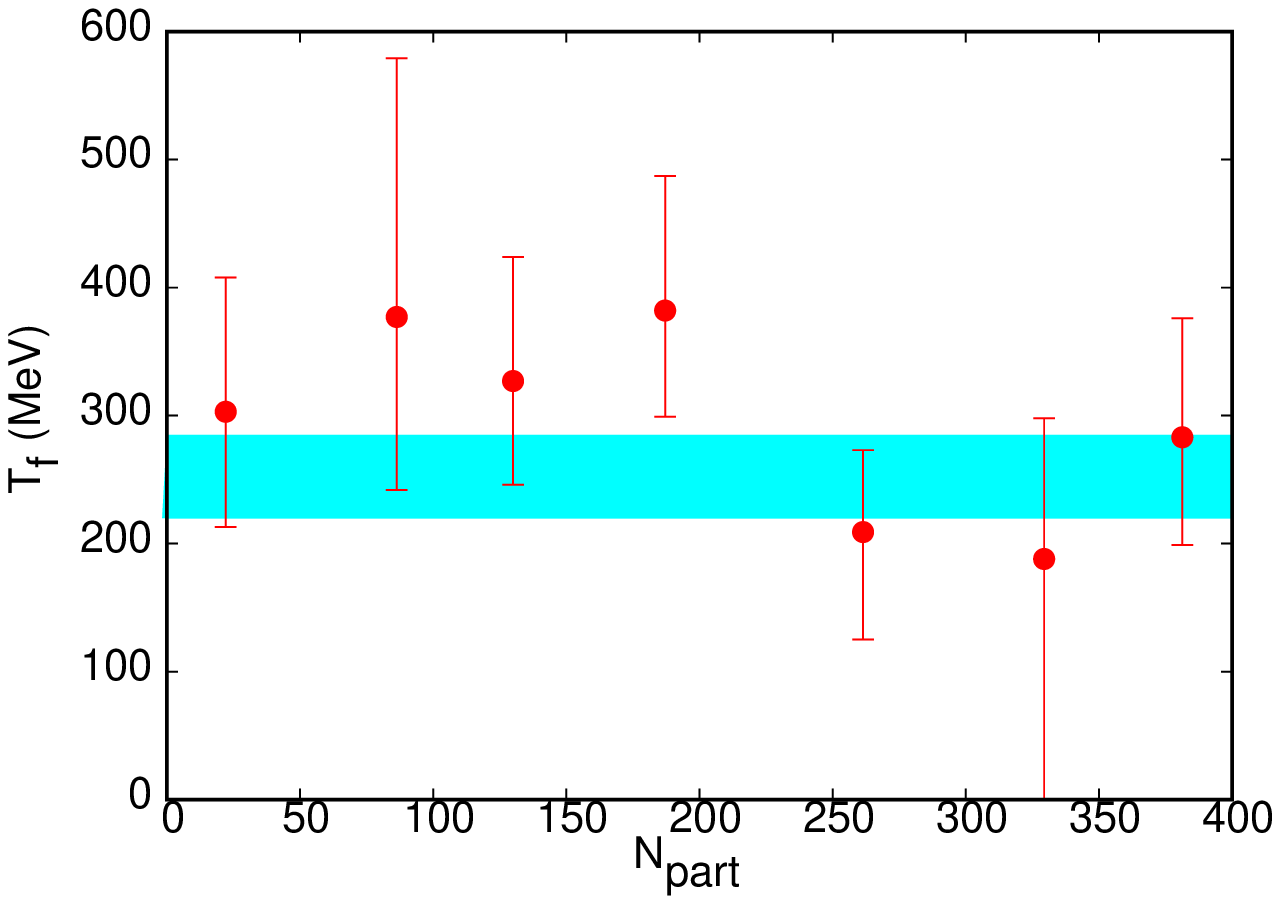}
\end{center}
\caption{The freezeout temperatures inferred from the CMS data on the
 centrality dependence of $Y[\Upsilon(2S)]$. The band shows the
 result quoted in \eqn{param}.  Within the large error bars shown it
 is not possible to say with certainty whether $T_f$ is larger for
 more peripheral collisions.}
\eef{centrality}


It is also possible to examine the centrality dependence of
the yields, although the errors are bound to be larger. First
estimates are presented in \fgn{centrality}. We have used the
double ratio reported in \cite{lhc-cms-upsilon} and the value of
$N_{pp}[\Upsilon(2S)]/N_{pp}[\Upsilon(1S)]$ reported there to get
$Y[\Upsilon(2S)]$ as a function of centrality. The errors in the double
ratio include the errors in this normalization, so adding them would
result in double counting. In fact, these errors should be removed from
the double ratio when we reconstruct $Y[\Upsilon(2S)]$ in different
centrality bins. Unfortunately this is not possible without access
to the data sample, so we just add in quadrature the statistical and
systematic uncertainties in the double ratio. The near constancy of $T_f$
could be a strong argument for thermalization, in the sense that the
initial state information is forgotten. 

However, it would be premature to come to this conclusion yet, in view
of the large errors. Here is one of the alternative scenarios which must
be examined with improved data. In peripheral collisions the separation
between the comovers and fireball may be much reduced, resulting in other
sources of suppression. This may result in the effective value of $T_f$
increasing in peripheral collisions. The values shown in the figure may
indicate such an effect, although it is hard to make any statistically
significant inference because of the large error bars. Nevertheless,
the fact that the four most peripheral bins lie on one side of the
mean, and the central bins lie largely on the other side could have
some significance. We believe that a re-analysis of the data along
these lines by the experimental collaboration should be able to give a
clearer picture.

We turn now to the preliminary data on {\jpsi} from the LHC.
$R_{\pbpb}[\jpsi]$ and its systematics have been reported by the
ALICE \cite{lhc-alice-jpsi}, ATLAS \cite{lhc-atlas-jpsi} and CMS
\cite{lhc-cms-jpsi} collaborations. However, the comparison of the
$\psi(2S)$ and {\jpsi} is yet to mature. CMS \cite{lhc-cms-psi} uses
the most central 0--20\% of the events to extract values
\beqa
\nonumber
   Y[\psi(2S)] &=& \frac{N_{\pbpb}[\psi(2S)]}{N_{\pbpb}[\jpsi]} \\
     &=& \begin{cases}
      0.024\pm0.008 & (|y|\le1.6), \cr
      0.105\pm0.02  & (1.5\le|y|\le2.4).
     \end{cases}
\eeqa{psi}
Both these values are boosted above the pp ratio, implying that the
fireball is a little richer in $\psi(2S)$ than an equivalent system
without re-interactions.  The central rapidity bin contains $p_\T$ between
6.5 and 30 GeV, whereas the peripheral bin has $p_\T$ between 3 and
30 GeV.  These give $T_f=149\pm14$ MeV at central rapidity but $255\pm25$
MeV at larger rapidity. Since both rapidity bins are well-separated from
the spectator rapidity, one does not expect this difference to be due
to comover suppression effects. It is more likely that the differences
arise from the different $p_\T$ acceptances. The higher $p_\T$ particles
escape more easily from the fireball and therefore may not thermalize
perfectly. If so, then the lower value of $T_f$ for higher $p_\T$
cutoff simply tells us that the charmonia which are sampled are not
in complete thermal equilibrium.  This is borne out by the results
reported by CMS on the $N_{part}$ dependence of the double ratio for
$|y|<1.6$. Our analysis seems to show that $T_f$ increases from about
185 MeV to 240 MeV as $N_{part}$ increases from 35 to 310. We believe
that the correct interpretation of this result is that the large $p_\T$
charmonium system goes from being less to more nearly thermalized in
more central collisions, \ie, as the fireball size increases. This
picture can also account for the fact that the systematics of the
double ratio $R_{\pbpb}[\psi(2S)]/R_{\pbpb}[\jpsi]$ reported by ALICE
\cite{lhc-alice-psi} is quite the opposite, and qualitatively similar
to the CMS observations for bottomonium \cite{lhc-cms-upsilon}. ALICE
has the advantage that it can detect {\jpsi} and $\psi(2S)$ with no
lower cutoff on $p_\T$. Once ALICE gives the pp normalization which
allows converting the double ratio to $Y[\psi(2S)]$, a genuine $T_f$
in the charmonium system can be extracted.

We have argued that sequential suppression of the $\Upsilon$ family of
mesons observed in the CMS experiment at LHC \cite{lhc-cms-upsilon} can
be interpreted as the system being in thermal equilibrium in the fireball
before freezing out. $R_{\pbpb}$ provides a simple diagnostic that the
physics of PbPb collisions is different from that in pp collisions, but
thermal behaviour is easier to analyze using the relative yields, $Y$,
defined in \eqn{cms-values}. The yields of $\Upsilon(nS)$ allow us to make
a first estimate of a freezeout temperature, $T_f=252\pm40$ MeV, which is
different from the freezeout temperatures seen in hadrons made of light
quarks. This is more or less constant as a function of the centrality,
with perhaps a very mild tendency to drop in the most central collisions.
This could mean a slightly earlier freezeout in peripheral collisions,
or a mild contamination by cold nuclear matter effects. This picture
predicts the yields of other bound states in the bottomonium sector. In
the charmonium sector the data is still preliminary. The absence of
complete data with low-$p_\T$ charmonia makes it impossible to extract
its $T_f$ now, although the situation may improve soon.  We have pointed
out various possible lines of investigation which probe such a simple
analysis more deeply.

Part of this work was carried out in the Workshop on High Energy Particle
Phenomenology (WHEPP XIII) held in Puri, India. We would like to thank
Saumen Datta and Rajiv Gavai for discussions, and Saumen Datta for comments
on the manuscript.

\end{document}